\documentclass[conference]{IEEEtran}

\ifCLASSINFOpdf
\else
\fi

\usepackage{graphicx}%
\usepackage{bm}%
\usepackage{amsfonts}
\usepackage{amssymb}
\usepackage{times}
\usepackage{subfigure}
\usepackage{enumitem}
\usepackage{latexsym,bm,amsmath,amssymb} 
\usepackage{CJK}
\usepackage{hhline}
\usepackage{cite}

\usepackage{multirow} 
\usepackage{xcolor}
\usepackage[linesnumbered,ruled,vlined]{algorithm2e}
\usepackage{ntheorem}
\usepackage{epstopdf}
\newtheorem{lemma}{\bf{Lemma}}
\newtheorem*{proof}{\it{Proof:}}

\usepackage{geometry}
\begin{document}

\title{Improving Channel Estimation Performance for Uplink OTFS Transmissions: Pilot Design based on \emph{A Posteriori} Cram$\Acute{\text{e}}$r-Rao Bound}

\vspace{-2mm}
\author{\IEEEauthorblockN{Mingcheng Nie\IEEEauthorrefmark{1},
Shuangyang Li\IEEEauthorrefmark{2}, and
Deepak Mishra\IEEEauthorrefmark{1}
}
\IEEEauthorblockA{\IEEEauthorrefmark{1}School of Electrical Engineering and Telecommunications, University of New South Wales, Sydney, NSW 2052, Australia\\
\IEEEauthorrefmark{2}School of Engineering, University of Western Australia, Perth, WA 6009, Australia\\
Emails: m.nie@student.unsw.edu.au, shuangyang.li@uwa.edu.au,  and d.mishra@unsw.edu.au }
}


\maketitle

\begin{abstract}
Orthogonal time frequency space (OTFS) has been widely acknowledged as a promising wireless technology for challenging transmission scenarios, including high-mobility channels. In this paper, we investigate the pilot design for the multi-user OTFS system based on the \emph{a priori} statistical channel state information (CSI), where the practical threshold-based estimation scheme is adopted. Specifically, we first derive the $\emph{a posteriori}$ Cram$\Acute{\text{e}}$r-Rao bound (PCRB) based on \emph{a priori} channel information for each user. According to our derivation, the PCRB only relates to the user's pilot signal-to-noise ratio (SNR) and the range of delay and Doppler shifts under the practical power-delay and power-Doppler profiles. Then, a pilot scheme is proposed to minimize the average PCRB of different users, where a closed-form global optimal  pilot power allocation is derived. Our numerical results verify the multi-user PCRB analysis. Also, we demonstrate an around 3 dB improvement in the average normalized-mean-square error (NMSE) by using the proposed pilot design in comparison to the conventional embedded pilot design under the same total pilot power.
\end{abstract}


%
\IEEEpeerreviewmaketitle
\vspace{-1mm}
\section{Introduction}
\vspace{-1mm}
One of the core objectives of next-generation wireless communications is to provide robust and reliable services in challenging scenarios, including high-mobility channels. For such channels, the current time-frequency domain (TF) modulation, i.e., orthogonal frequency division modulation (OFDM), could suffer from severe Doppler effects, damaging the orthogonality between subcarriers. In contrast, orthogonal time frequency space (OTFS) is a promising delay-Doppler (DD) domain modulation technology capable of tackling such channels~\cite{hadani2017orthogonal,wei2021orthogonal}. Specifically, the OTFS modulation multiplexes the data symbols in the DD domain and spreads each symbol to the whole TF domain, where all DD domain symbols will theoretically experience the same TF domain doubly-selective channel. Consequently, the effective time-invariant channel response in the DD domain can be sparsely represented by only a few parameters, facilitating simple channel estimation~\cite{wei2022off} and transceiver designs~\cite{li2021cross,li2021hybrid}. 
\vspace{-0.5mm}
\subsection{State-of-the-Art and Motivation}
\vspace{-0.5mm}
To perform reliable data detection, accurate channel state information (CSI) is crucial. Thus, efficient channel estimation schemes with low complexity in the DD domain are highly desirable. An embedded pilot scheme was proposed in~\cite{raviteja2019embedded}, where a sufficiently large guard space was adopted to eliminate the overlapping between pilot and information symbols at the receiver side. The channel can be simply estimated by comparing the received symbols in the guard space with a threshold, where the estimation accuracy depends on the pilot and channel power due to the convolution property of DD domain channels. Furthermore, the authors in~\cite{yuan2021data} proposed a superimposed pilot scheme, which removed the guard space for improving spectral efficiency. Since the interference between pilot symbol and data symbols is no longer avoidable without guard space, iterative processing among threshold-based channel estimation, interference cancellation, and data detection was also proposed in~\cite{yuan2021data}, where the threshold is modified per iteration based on detected data to refine the channel estimation. The numerical results in~\cite{raviteja2019embedded,yuan2021data} demonstrated that a higher pilot power can lead to a better channel estimation performance. Although this conclusion is not surprising, there is no quantitive research on the impact of pilot power on the channel estimation performance for OTFS transmissions available in the literature, to the best knowledge of the authors.

The Cram$\Acute{\text{e}}$r-Rao bound (CRB) is a widely applied mathematical tool characterizing optimal channel estimation performance, and its application in OTFS transmissions have also been studied in~\cite{gaudio2020effectiveness,liu2021message}. For example, the CRB for the joint sensing and communication using OTFS signals was first derived in~\cite{gaudio2020effectiveness}, which can be achieved by using the proposed approximated maximum likelihood (ML) estimator. The CRB of channel coefficient and fractional Doppler estimation was derived in~\cite{liu2021message} as a benchmark to evaluate the proposed estimation algorithm. However, we observe that the actual CSI is needed in the derivation of the CRB~\cite{gaudio2020effectiveness,liu2021message}, and consequently CRB can only provide limited insights for practical designs, where CSI at the transmitter is typically unknown~\cite{MishraTSP}. It should be highlighted that the expected mean-square-estimation-error (MSE) performance under optimal estimation is bounded by the $\emph{a posteriori}$ Cram$\Acute{\text{e}}$r-Rao bound (PCRB), which relies on the \emph{a priori} channel statistical information instead of the actual CSI for each channel realization. More importantly, the impact of the pilot design can also be reflected in PCRB as we will show later, which provides valuable insights for practical system implementations. 
\vspace{-2mm}
\subsection{Contributions and Notations}
\vspace{-1mm}
The main contributions are listed as follows
\begin{itemize}[leftmargin=*] 
    \item Using \emph{a priori} channel statistical information, we derive the closed-form PCRB for uplink multi-user OTFS with practical power-delay and power-Doppler profiles, where the PCRB remains independent from delay and Doppler indices.
    \item A pilot design scheme is proposed to improve the channel estimation accuracy on the system level by minimizing the average PCRB of different users, where we show the pilot design can be reduced to a power allocation problem, whose optimal solution is then derived.
    \item Numerical results verify the multi-user average normalized-mean-square error (NMSE) improvement compared to the conventional embedded pilot design for a given power budget. 
\end{itemize}
\emph{Notations:} $(\cdot)^{*}$, $\left\| {\cdot} \right\|$, $| {\cdot} |$, $(\cdot)^{-1}$, and $(\cdot)^{\rm{H}}$ denote conjugate, Euclidean norm, absolute, inverse, and Hermitian operations, respectively; $\mathbb{E}_{A}[\cdot]$ denotes expectation with respect to $A$; $\widehat{A}$ represents the estimate of $A$; $\Re{\{A\}}$ and $\Im{\{A\}}$ denote the real and imaginary part of complex $A$, respectively.
\vspace{-1mm}
\section{System Model} 
\vspace{-1mm}
\subsection{Multi-user OTFS Transmission}
\vspace{-0.5mm}
In this section, we present an uplink multi-user OTFS model with $K_u$ users, where each user follows the transmission structure in~\cite{raviteja2018interference} with a single antenna, practical rectangular pulses, and integer delay and Doppler indices. Specifically, we consider a two-dimensional grid $\Gamma$ in the DD domain as $\Gamma = \{(\frac{l}{M\Delta f},\frac{k}{NT}), l=0, \cdots, M-1, k=0, \cdots, N-1\}$, where $\frac{1}{M\Delta f}$ and $\frac{1}{NT}$ are the sampling intervals along delay and Doppler axis, respectively. Here $M$ denotes the number of delay bins$/$number of subcarriers, $N$ denotes the number of Doppler bins$/$number of time slots, $\Delta f$ is the subcarrier spacing, and $T$ is the time slots duration~\cite{hadani2017orthogonal}.
\begin{figure}
    \centering
    \includegraphics[scale=0.5]{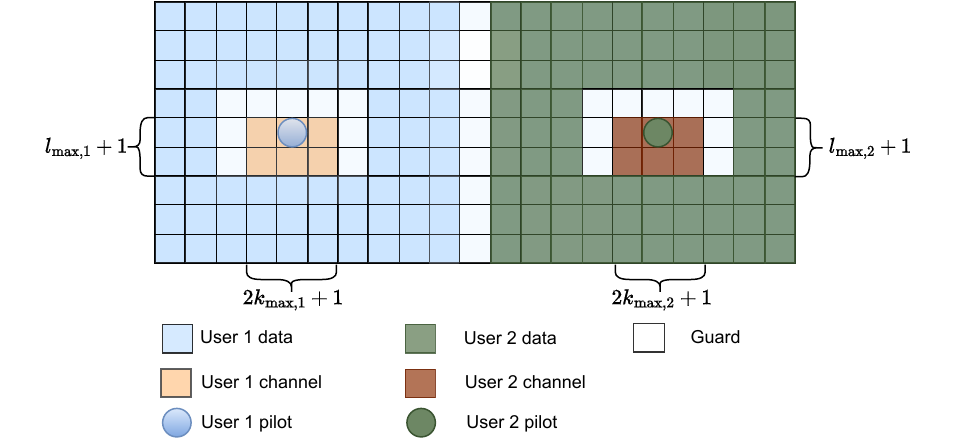}
    \vspace{-4mm} 
    \caption{Received signals of two users }
    \label{fig:system_model}
    \vspace{-5mm} 
\end{figure}

We adopt the DD domain pilot structure in~\cite{raviteja2019embedded} for each user as shown in Fig. \ref{fig:system_model}. We assume that the information symbols of different users are sufficiently separated by the guard space. Let $x_u[k,l]$ be the $(k,l)$-th DD domain symbol of the $u$-th user that is transformed to the TF domain by the inverse symplectic finite Fourier transform (ISFFT)~\cite{hadani2017orthogonal}, i.e.,
\begin{eqnarray} 
X_u[n,m]=&\hspace{-2mm}\frac{1}{\sqrt{NM}}\sum\limits_{k=0}^{N-1}\sum\limits_{l=0}^{M-1}x_u[k,l]e^{j2\pi(\frac{nk}{N}-\frac{ml}{M})},
\label{1}
\end{eqnarray}
where $0\le m\le M-1$ and $0\le n\le N-1$. Then, applying the Heisenberg transform yields the following~\cite{hadani2017orthogonal}
\begin{eqnarray} 
s_u(t)=&\hspace{-2mm}\sum\limits_{n=0}^{N-1}\sum\limits_{m=0}^{M-1}X_u[n,m]e^{j2\pi m\Delta f(t-nT)}g_{\rm{tx}}(t-nT)\label{2},
\end{eqnarray}
where $g_{\rm{tx}}(t-nT)$ is the rectangular transmitting pulse~\cite{raviteja2018interference}. The received signal of the $u$-th user is given as~\cite{hadani2017orthogonal}
\begin{eqnarray}
r_u(t)=&\hspace{-2mm}\int_\tau\int_\nu h_u(\tau,\nu)e^{j2\pi\nu(t-\tau)}s_u(t-\tau)d\nu d\tau\label{3},
\end{eqnarray}
where we ignore the noise term for simplicity. In~(\ref{3}), $h_u(\tau,\nu)$ is the complex baseband doubly-selective channel response of the $u$-th user and is given in the sparse representation as~\cite{jakes1994microwave}
\begin{eqnarray}
    h_u(\tau,\nu) =&\hspace{-2mm} \sum\limits_{p=1}^{P} h_{p,u} \delta(\tau -\tau_{p,u}) \delta(\nu -\nu_{p,u})\label{4},
\end{eqnarray} 
where $h_{p,u}$, $\tau_{p,u}=\frac{l_{{p,u}}}{M\Delta f}$, $\nu_{p,u}=\frac{k_{{p,u}}}{NT}$, $l_{{p,u}}$, and $k_{{p,u}}$ are the complex channel coefficient, delay shift, Doppler shift, delay index, and Doppler index of the $p$-th path of the $u$-th user, respectively. We assume that each user has $P$ independent resolvable channel paths. After sampling the output of the rectangular matched filter, the symplectic finite Fourier transform (SFFT) is applied to transform the TF domain signal $Y_u[n,m]$ into the DD domain as~\cite{hadani2017orthogonal}
\begin{equation}
y_u[k,l]=\frac{1}{\sqrt{NM}}\sum_{n=0}^{N-1}\sum_{m=0}^{M-1}Y_u[n,\ m]e^{-j2\pi \left(\frac{nk}{N}-\frac{ml}{M}\right)}\label{5}.
\end{equation}

According to~\cite{raviteja2018interference}, we can obtain the input-output relationship for multi-user OTFS transmissions based on $\eqref{1}-\eqref{5}$ with rectangular pulses  in the DD domain  by 
\begin{align}
    y[k,l]=z[k,l]+w[k,l],\label{input-output}
\end{align} 
$y[k,l]$ denotes the received signal at the base station (BS), $w[k,l]$ is the equivalent additive complex Gaussian noise with power spectrum density (PSD) $N_0$, and $z[k,l]$ is
\begin{eqnarray}\label{z}
z[k,l]=&\hspace{-4mm}\sum\limits_{u=1}^{K_u}\hspace{-0.5mm}\sum\limits_{p=1}^{P}\!h_{p,u}e^{j2\pi\frac{(l-l_{{p,u}})k_{{p,u}}}{MN}}\!x_u[[k\!-\!k_{{p,u}}]_{N},[l\!-\!l_{{p,u}}]_{M}].\hspace{-2mm}\nonumber\\
\end{eqnarray}\vspace{-7mm}

 Note that the channel can be estimated by setting a threshold comparison, i.e., $y[k,l]=0,\forall |y[k,l]|<\beta$~\cite{raviteja2019embedded}, then remove the pilot and correct the phase as shown in~\cite{raviteja2018interference} for the practical rectangular pulse case.

\subsection{A Priori Distribution of Channel Parameters}
Without loss of generality, we assume $l_{p,u}$ and $k_{p,u}$ are uniformly obtained from $[l_{\min,u},l_{\max,u}]$ and $[-k_{\max,u},k_{\max,u}]$, respectively, where $l_{\min,u}$, $l_{\max,u}$, and $k_{\max,u}$ represent the minimum delay, maximum delay, and maximum Doppler indices of the $u$-th user, respectively. The \emph{a priori} probability density functions (PDFs) of $l_{p,u}$ and $k_{p,u}$ are given by~\cite{walpole1993probability}
\begin{align}
    p(l_{p,u})&=
    \begin{cases}
    \frac{1}{l_{\max,u}-l_{\min,u}},\quad &l_{\min,u} \le l_{p,u} \le l_{\max,u},\\
    \quad \quad 0, \qquad\quad\  &\qquad\qquad\qquad\quad \ \,\rm{else},
    \end{cases}\\
    p(k_{p,u})&=
    \begin{cases}
    \frac{1}{2k_{\max,u}}, \quad &-k_{\max,u} \le k_{p,u} \le k_{\max,u},\\
    \quad  0, \quad  &\qquad\qquad\qquad\qquad\quad \, \rm{else},
    \end{cases}
\end{align}
and the joint PDF of delay and Doppler indices of the $u$-th user is obtained by $p(\mathbf{l}_u)=\prod_{p=1}^{P}p(l_{p,u})$ and $p(\mathbf{k}_u)=\prod_{p=1}^{P}p(k_{p,u})$ due to the assumption of under-spread wide-sense stationary uncorrelated scattering (WSSUS) channels. Assume that the channel coefficient of the $u$-th user's $p$-th path follows $h_{p,u}\sim\mathcal{CN}(0,\gamma_{p,u})$, where $\gamma_{p,u}$ is the variance of the corresponding channel coefficient that follows exponential power-delay and uniform power-Doppler profiles
as~\cite{hlawatsch2011wireless}
\begin{align}
    \gamma_{p,u}=\frac{e^{-{l_{p,u}}}}{2k_{\max,u}}.\label{powerprofile}
\end{align}
We note that the U-shape Jake's power-Doppler profile is also widely considered. However, it is not compiled with PCRB derivation, where the PDFs are required to be square-summable~\cite{dauwels2005computing}. It should be noted that our proposed derivation can be extended to Jake's power-Doppler profile with a truncation on the corresponding PDF~\cite{dauwels2005computing}. Following~(\ref{powerprofile}), the \emph{a priori} PDF of $h_{p,u}$ is given by
\begin{align}\label{p(h)}
    p(h_{p,u})=\frac{1}{{\pi \gamma_{p,u}}}\exp\left(-\frac{|h_{p,u}|^2}{\gamma_{p,u}}\right),
\end{align}
where $|\cdot|$ is the absolute value. Thus, the joint PDF of channel coefficient of $u$-th user is obtained by $p(\mathbf{h}_u)=\prod_{p=1}^{P}p(h_{p,u}).$ due to independency assumption for resolvable channel paths.

\section{A Posteriori Cram$\Acute{\text{e}}$r-Rao Bound Analysis}
\subsection{A Posteriori Distribution of Channel Parameters}
First, we define a vector that contains the channel parameters in~(\ref{4}). Note the channel coefficient is complex, and we decompose it into real and imaginary parts. Thus, we define a length-$4PK_u$ channel parameter vector $\boldsymbol{\theta}=[\mathbf{h},\mathbf{l},\mathbf{k}],$
where $\mathbf{h}=[\mathbf{h}_1,\cdots,\mathbf{h}_u,\cdots,\mathbf{h}_{K_u}]$ is a length-$2PK_u$ real vector contains the real and imaginary parts of channel coefficients, $\mathbf{l}=[\mathbf{l}_1,\cdots,\mathbf{l}_u,\cdots,\mathbf{l}_{K_u}]$ is a length-$PK_u$ vector contains the delay indices, and $\mathbf{k}=[\mathbf{k}_1,\cdots,\mathbf{k}_u,\cdots,\mathbf{k}_{K_u}]$ is a length-$PK_u$ vector contains the Doppler indices of all users. Specifically, $\mathbf{h}_u$, $\mathbf{l}_u$, and $\mathbf{k}_u$ for each user are given as
\begin{align}
\mathbf{h}_u =[&\Re(h_{1,u}),\cdots,\Re(h_{P,u}),\Im(h_{1,u}),\cdots,\Im(h_{P,u})],\\
\mathbf{l}_u =[&l_{1,u},\cdots,l_{P,u}], \rm{and} \ \mathbf{k}_u =[k_{1,u},\cdots,k_{P,u}],
\end{align}
where $\Re(h_{p,u})$ and $\Im(h_{p,u})$ represent the real and imaginary part of $h_{p,u}$. 
Based on the Bayesian theorem and the \emph{a priori} information $p(\boldsymbol{\theta})$, the \emph{a posteriori} PDF of $\boldsymbol{\theta}$ for given $\mathbf{y}$ is~\cite{kay1993fundamentals}
\begin{align}
    p(\boldsymbol{\theta}|\mathbf{y})&=\frac{p(\boldsymbol{\theta},\mathbf{y})}{p(\mathbf{y})}\propto p(\mathbf{y}|\boldsymbol{\theta})p(\boldsymbol{\theta}),
\end{align}
where $\mathbf{y}$ is the receive vector by stacking the corresponding symbols in~(\ref{input-output}). According to the power profiles we adopt in~(\ref{powerprofile}), we can obtain
\begin{align}
         p(\boldsymbol{\theta}|\mathbf{y})\propto p(\mathbf{y}|\boldsymbol{\theta})p(\mathbf{h}|\mathbf{k},\mathbf{l})p(\mathbf{k})p(\mathbf{l}).\label{postPDF}
\end{align}

\subsection{A {Posterior} Cram$\Acute{\text{e}}$r-Rao Bound Derivation}

According to \cite{dauwels2005computing}, the expected MSE matrix of $\boldsymbol{\theta}$ is bounded by the inverse of the \emph{a posteriori} Fisher information matrix (FIM) $\mathbf{J}$ as $\mathbb{E}\left[(\widehat{\boldsymbol{\theta}}-\boldsymbol{\theta})(\widehat{\boldsymbol{\theta}}-\boldsymbol{\theta})^{T}\right]\succeq\mathbf{J}^{-1}\triangleq\boldsymbol{\alpha}$. Thus, we define the PCRBs of channel coefficient, delay, and Doppler indices by
\begin{align}
    &\mathbb{E}\left[||\widehat{\mathbf{h}}-\mathbf{h}||^2 \right]\ge \sum_{i=1}^{2PK_u}\boldsymbol{\alpha}_{(i,i)}\triangleq {\rm{PCRB}}(\mathbf{h})\label{PCRB_h},\\
    &\mathbb{E}\left[||\widehat{\mathbf{l}}-\mathbf{l}||^2 \right]\ge \sum_{i=2PK_u+1}^{3PK_u}\boldsymbol{\alpha}_{(i,i)}\triangleq {\rm{PCRB}}(\mathbf{l})\label{PCRB_l},\\
    &\mathbb{E}\left[||\widehat{\mathbf{k}}-\mathbf{k}||^2 \right]\ge \sum_{i=3PK_u+1}^{4PK_u}\boldsymbol{\alpha}_{(i,i)}\triangleq {\rm{PCRB}}(\mathbf{k})\label{PCRB_k},
\end{align}
where $\boldsymbol{\alpha}_{(i,i)}$ is the $i$-th diagonal entry of $\boldsymbol{\alpha}$ and the \emph{a posteriori} FIM $\mathbf{J}$ can be obtained based on~(\ref{postPDF}) by~\cite{dauwels2005computing} 
\begin{align}
    \mathbf{J}^{(i,j)}&=-\mathbb{E}_{ \boldsymbol{\theta} \mathbf{y}}\left[\frac{\partial^2 \ln p(\boldsymbol{\theta}|\mathbf{z},\mathbf{y})}{\partial\theta_i\partial\theta_j}\right],\nonumber\\
    &=-\mathbb{E}_{ \boldsymbol{\theta} \mathbf{y}}\left[\frac{\partial^2 \ln  {p(\mathbf{y}|\boldsymbol{\theta},\mathbf{z})} }{\partial\theta_i\partial\theta_j}\right]-\mathbb{E}_{ \boldsymbol{\theta}}\left[\frac{\partial^2\ln p(\mathbf{h}|\mathbf{k},\mathbf{l})}{\partial\theta_i\partial\theta_j}\right],\nonumber\\
    &=\mathbf{J}_{r}^{(i,j)}+\mathbf{J}_{p}^{(i,j)},\label{postFIM}
\end{align}\vspace{-0.05mm}
where $\mathbf{J}^{(i,j)}$ is the $(i,j)$-th entry of $\mathbf{J}$ and we omit the data symbols in deriviations. Compared with conventional CRB's FIM $\mathbf{F}=-\mathbb{E}_{ \mathbf{y}}\left[\frac{\partial^2 \ln  {p(\mathbf{y}|\boldsymbol{\theta},\mathbf{z})} }{\partial\theta_i\partial\theta_j}\right]$, the \emph{a priori} information matrix $\mathbf{J}_{p}$ consist of the statistical information for the estimation, where the expected channel power in (\ref{powerprofile}) is reflected. Note that the operation $\mathbb{E}_{\boldsymbol{\theta}}[\cdot]$ ensures the PCRB is independent of the channel coefficient, delay and Doppler indices. Thus, the conditional logarithm PDF of $\mathbf{y}$ is given as follows 
\begin{align}
    \ln{p(\mathbf{y}|\boldsymbol{\theta},\mathbf{z})}=-MN\ln{\left(\pi N_0\right)}-\frac{1}{N_0}\sum_{k,l} \lvert y[k,l]-z[k,l]\rvert^2.\label{condition_y}
\end{align}
According to~(\ref{postFIM}) and (\ref{condition_y}), we can derive the following lemma.
\begin{lemma}
The $(i,j)$-th entry of $4PK_u\times 4PK_u$ FIM $\mathbf{J}_r$ is obtained by
\end{lemma}
\begin{align}
    \mathbf{J}_r^{(i,j)}=\frac{2}{N_0}\mathbb{E}_{\boldsymbol{\theta}}\left[\Re\left\{\sum_{k,l}\left( \frac{\partial z[k,l]}{\partial \theta_i}\right)^*\left(\frac{\partial z[k,l]}{\partial \theta_j}\right)\right\}\right],\label{J_r}
\end{align}
\begin{proof}
\rm{The proof is given in Appendix A.}
\end{proof}
By defining $x_u[k',l']\triangleq x_u[[k-k_{p,u}]_{N},[l-l_{p,u}]_{M}]$, the partial derivatives in~(\ref{J_r}) are given by
\begin{align}
\frac{\partial z[k,l]}{\partial  \Re{(h_{p,u})}} &= e^{j2\pi\frac{(l-l_{p,u})k_{p,u}}{MN}}x_u[k',l'],\label{h_R}\\
\frac{\partial z[k,l]}{\partial \Im(h_{p,u})} &=j e^{j2\pi\frac{(l-l_{p,u})k_{p,u}}{MN}}x_u[k',l'],\label{h_I}\\
\frac{\partial z[k,l]}{\partial l_{p,u}} &= -h_{p,u} \frac{j2\pi k_{p,u}}{MN}e^{j2\pi\frac{(l-l_{p,u})k_{p,u}}{MN}}x_u[k',l'], \rm{and}\label{l}\\
\frac{\partial z[k,l]}{\partial k_{p,u}} &= h_{p,u}\frac{j2\pi(l-l_{p,u})}{MN}e^{j2\pi\frac{(l-l_{p,u})k_{p,u}}{MN}}x_u[k',l']\label{k},
\end{align}
respectively.
By substituting the partial derivatives~(\ref{h_R})-(\ref{k}) into~(\ref{J_r}) and under the assumption of a single pilot for each user, we can observe that $\mathbf{J}_r$ is a diagonal matrix (see Appendix A). Thus, the $i$-th diagonal entry of $\mathbf{J}_r$ satisfies
\begin{align}
    \mathbf{J}_r^{(i,i)}=
    \begin{cases}
    \frac{2\mathbf{x}_u^H\mathbf{x}_u}{N0}, &\qquad 1\le i \le 2PK_u,\\
    \frac{2\mathbf{x}_u^H\mathbf{x}_u}{N0}{A}, &2PK_u< i \le 3PK_u,\\
    \frac{2\mathbf{x}_u^H\mathbf{x}_u}{N0}{B}, &3PK_u< i \le 4PK_u,\\
    0, &\qquad\qquad\qquad\ \ \rm{else},
    \end{cases}\label{J_s}
\end{align}
where $\mathbf{x}_u$ is obtained by stacking $x_u[k',l']$, ${A}=\mathbb{E}_{\boldsymbol{\theta}}\left[\left(|h_{p,u}|\frac{2\pi k_{p,u}}{MN}\right)^2\right]$, ${B}=\mathbb{E}_{\boldsymbol{\theta}}\left[\left(|h_{p,u}|\frac{2\pi (l''-l_{p,u})}{MN}\right)^2\right]$, $l''=l_{p,u}+l_{\tau,u}$, and $l_{\tau,u}$ is the pilot delay index of $u$-th user. Based on~(\ref{p(h)}) and~(\ref{postFIM}), we have the following lemma.
\begin{lemma}
\rm{The FIM $\mathbf{J}_p$ is also diagonal and the $i$-th diagonal entry is obtained as follow}
\end{lemma}
\begin{align}
    \mathbf{J}_p^{(i,i)}=
    \begin{cases}
    \mathbb{E}_{\boldsymbol{\theta}}\left[\frac{2}{\gamma_{p,u}}\right], &\quad\quad 1\le i \le 2PK_u,\\
    \mathbb{E}_{\boldsymbol{\theta}}\left[\left(\frac{1}{\gamma_{p,u}}\frac{\partial \gamma_{p,u}}{\partial l_{p,u}}\right)^2\right], &2PK_u< i \le 3PK_u,\\
    \mathbb{E}_{\boldsymbol{\theta}}\left[\left(\frac{1}{\gamma_{p,u}}\frac{\partial \gamma_{p,u}}{\partial k_{p,u}}\right)^2\right], &3PK_u< i \le 4PK_u,\\
    0, &\qquad\qquad\qquad\;\ \rm{else},
    \end{cases}\label{J_p}
\end{align}
\begin{proof}
\rm{The proof is given in Appendix B.}
\end{proof}

Now, we can observe from~(\ref{PCRB_h})-(\ref{postFIM}) that the PCRB analysis relates to both the \emph{a priori} information and the likelihood function $p(\mathbf{y}|\boldsymbol{\theta},\mathbf{z})$, whereas the conventional CRB only relates to the likelihood function~\cite{dauwels2005computing}. As a result, the PCRB contains important information of the statistical channel characteristics rather than the temporal channel information that is only valid in a particular transmission, which can provide clear insights on the average MSE ergodically. More importantly, the actual CSI for each transmission is hard to obtain at the transmitter side, while the statistical channel information in~(\ref{J_p}) is generally available in the system design. Therefore, PCRB provides more implementation significance for communication design. Our pilot design will be investigated in Section~\ref{sec:power allocation}.

Next, we present the close form PCRB of the channel coefficients in~(\ref{PCRB_h}) based on~(\ref{J_s}) and (\ref{J_p}) as follows
\begin{align}
    &{\rm{PCRB}}(\mathbf{h})=\sum_{u=1}^{K_u}\sum_{p=1}^{P}\left({\frac{\mathbf{x}^{H}_u\mathbf{x}_u}{N_0}}+\mathbb{E}_{\boldsymbol{\theta}}\left[\frac{1}{\gamma_{p,u}}\right]\right)^{-1}\label{GeneralClosePCRB},\\
    &\overset{\Omega}{=}\sum_{u=1}^{K_u} P\left({\frac{\mathbf{x}^{H}_u\mathbf{x}_u}{N_0}}+\underbrace{\frac{2 k_{\max,u}\left({e}^{l_{\max,u}}-{e}^{l_{\min,u}}\right)}{l_{\max,u}-l_{\min,u}}}_{\Phi_u}\right)^{-1},\label{closePCRB}
\end{align}
where $\Omega$ denotes substituting the exponential power-delay and uniform power-Doppler profiles in~(\ref{powerprofile}), $N_0$ is the noise PSD, and $\Phi_u$ denotes the $u$-th user's \emph{a priori} channel statistical information that only relates to the range of the delay and Doppler indices. The result in~(\ref{GeneralClosePCRB}) demonstrates that the PCRB of the multi-user is the accumulation of each user's PCRB and the PCRB of a multi-path channel is the accumulation of each path's PCRB, which can be validated by the fact that different users and paths of the channel are independent. Moreover, we can observe from (\ref{GeneralClosePCRB}) that the PCRB of each user relates to the corresponding pilot SNR and the channel statistic characteristic, i.e., the \emph{expectation} of the inverse of the channel variance. Thanks to the expectation property of PCRB, we show that bounds in (\ref{closePCRB}) only depend on the range of delay and Doppler indices even under the practical exponential power-delay and uniform power-Doppler profiles, which can be known at the transmitter side. Note that the detailed close form PCRBs of delay and Doppler indices can be obtained in a similar approach by calculating the corresponding components in~(\ref{J_s}) and (\ref{J_p}).

\vspace{-0.5mm}
\section{Pilot Design for Multi-user Transmissions Based on PCRB}\label{sec:power allocation}

In this section, we investigate the pilot design for multi-user based on the derived PCRB. It should be noted that the classical threshold-based channel estimation performs reasonably well at high SNRs. Therefore, we expect that optimizing the PCRB can further improve the estimation performance of the threshold-based channel estimation. Based on the discussion in the previous section, we propose to design pilots based on the PCRB of the channel coefficient. Note that the delay and Doppler have sufficiently high resolution with large $M$ and $N$ values, and consequently they are unlikely to introduce estimation error in practical systems at high SNRs compared to the channel coefficient. Based on~(\ref{closePCRB}), the corresponding pilot design problem can be fully represented by its pilot power. Thus, we propose to design the pilot by minimizing the average normalized PCRB (NPCRB) of channel coefficient estimation in~(\ref{GeneralClosePCRB}) and formulate it as
\begin{align}
\mathcal{O}:&\min_{\mathcal{P}_u}\;{\rm{NPCRB}}(\mathbf{h})=\sum_{u=1}^{K_u} N_u P\left(\mathcal{P}_u+\Phi_u\right)^{-1},\nonumber\\
\rm{subject}\; \rm{to:}&\qquad\qquad\quad\;\  \mathcal{P}=\sum_{u=1}^{K_u}\mathcal{P}_u,
\end{align}
where $N_u=\mathbb{E}\left[\frac{1}{||\mathbf{h}_u||^2}\right]$
is the normalization factor of $u$-th user's PCRB, $P$ denotes the number of channel paths, $\mathcal{P}$ is the total pilot power of the cell, $\mathcal{P}_u\triangleq\mathbf{x}_u^H\mathbf{x}_u, \forall u=1,\cdots,K_u$, is the pilot power of the $u$-th user,  and $\Phi_u$ is shown in~(\ref{closePCRB}).
Next, considering $\lambda$ as a Lagrangian multiplier for problem $\mathcal{O}$, we can write the underlying Lagrangian as follows
\begin{align}
    \mathcal{L}(\lambda,\mathcal{P}_1,\cdots,\mathcal{P}_u)=P\sum_{u=1}^{K_u} N_u\left(\mathcal{P}_u+\Phi_u\right)^{-1}+\lambda \sum_{u=1}^{K_u}\mathcal{P}_u.
\end{align}
Then, we can find the optimal power allocation for $\mathcal{O}$ by solving the subgradient Karush–Kuhn–Tucker (KKT) conditions by $\frac{\partial \mathcal{L}}{\partial \mathcal{P}_u}=0$, yields
\begin{align}
    \mathcal{P}_u=\sqrt{\frac{PN_u}{\lambda}}-\Phi_u.\label{Lag1}
\end{align}
After substituting~(\ref{Lag1}) into the constraint, we can obtain the optimal value of $\lambda$ as follows
\begin{align}
    \lambda=P\left(\frac{\sum_{u=1}^{K_u}\sqrt{N_u}}{\mathcal{P}+\sum_{u=1}^{K_u}\Phi_u}\right)^2,\label{lambda}
\end{align}
where $\lambda>0$. Finally, we can obtain the global optimal power allocation for each user by substituting~(\ref{lambda}) into~(\ref{Lag1}) as follows
\begin{align}
    \mathcal{P}_u=\frac{\sqrt{N_u}(\mathcal{P}+\sum_{u=1}^{K_u}\Phi_u)}{\sum_{u=1}^{K_u}\sqrt{N_u}}-\Phi_u.\label{optimal_power}
\end{align}
We observe from~(\ref{optimal_power}) that a small range of the delay and Doppler indices will lead to a small $\Phi_u$ and $N_u$ resulting in a small pilot power $\mathcal{P}_u$. This is because a small range of the delay and Doppler indices will statistically lead to a high average channel power according to the power profile~(\ref{powerprofile}), and therefore it does not need very high pilot power to have a good estimation performance.

\section{Numerical Results}\label{sec:numerical}
In this section, we present the normalized mean square error (NMSE) results for different users in terms of pilot SNR. Without other states, we set $M=32$, $N=32$, $P=4$, and $K_u=4$. Without loss of generality, we fix the noise PSD as $N_0=1$, such that the pilot $\text{SNR}_p$ shares the same value as the pilot power. The integer delay and Doppler indices are uniformly generated from $[l_{\min},l_{\max}]$ and $[-k_{\max},k_{\max} ]$, respectively. The channel coefficients are generated randomly from the exponential power-delay and uniform power-Doppler profiles based on (\ref{powerprofile}). We assume all users' power profiles are jointly normalized from the view of the receiver side. The delay and Doppler range of each user are shown in Table \ref{table:parameter}.
\begin{table}[htbp]
\caption{Multiuser Parameters}
\centering\vspace{-1.5mm}
\begin{tabular}{|c|c|c|c|}
\hline
User~&~$l_{\min}$~&~$l_{\max}$~&~$k_{\max}$\\
\hline
User1~&~0&~4~&~3\\
\hline
User2~&~6&~10~&~3\\
\hline
User3~&~0~&~4~&~7\\
\hline
User4~&~6&~10~&~7 \\
\hline
\end{tabular}
\vspace{-1mm}
\label{table:parameter}
\end{table}

\begin{figure}
\centering
\includegraphics[width=2.8in]{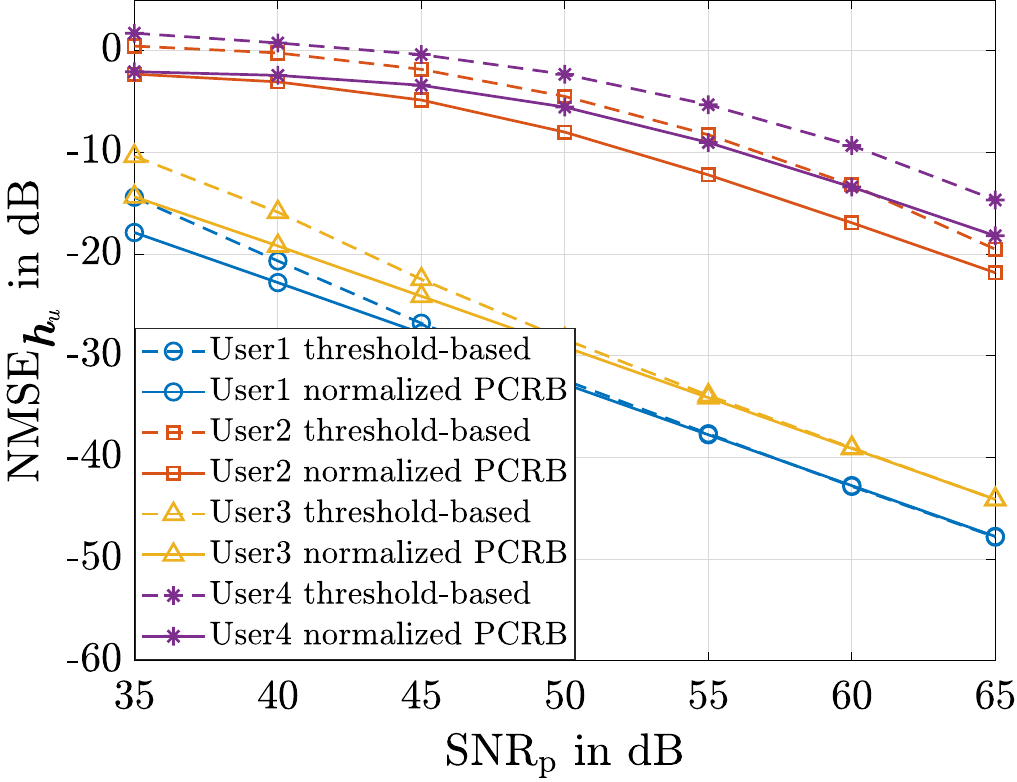}
    \vspace{-4mm} 
\caption{NMSE and normalized PCRB comparison for four users individually}
\label{fig:NMSE}
\end{figure}

We first demonstrate the PCRB and NMSE performance for different users individually in terms of pilot $\text{SNR}_p$ in Fig.~\ref{fig:NMSE}. The solid lines denote the normalized PCRB, and the dashed lines denote the corresponding threshold-based channel matrix estimation NMSE, i.e., ${\rm{NMSE}}_{\bm{h}_u}=\frac{||\hat{\bm{h}}_u-\bm{h}_u||^2}{||\bm{h}_u||^2}$, where $\bm{h}_u$ is the actual channel in~(\ref{4}) and $\hat{\bm{h}}_u$ denotes the estimated channel. We can observe the threshold-based ${\rm{NMSE}}_{\bm{h}_u}$ of users coincide with the corresponding normalized PCRB at high pilot SNRs, where the user who obtains better statistical channel condition will coincide with corresponding PCRB at lower SNR as expected. Note that user 2 and user 4 will coincide with the normalized PCRB at much higher pilot SNRs but omitted due to space limitations. The users with different delay and Doppler ranges have different estimation performances, which shows that pilot power allocation among multi-users to improve the overall estimation performances is possible. Moreover, with the user 1 result in hand, we can observe that user 3 has the same delay range but a larger Doppler range than user 1 and obtains a better estimation performance than user 2, who has the same Doppler range, but larger delay indices. This indicates that the user's estimation performance is more sensitive to the delay range rather than the Doppler range by considering the exponential delay and uniform Doppler power profiles. Note that user 4 has the largest delay and Doppler indices and obtains the worst estimation performance as expected.

\begin{figure}
\centering
\includegraphics[width=2.4in]{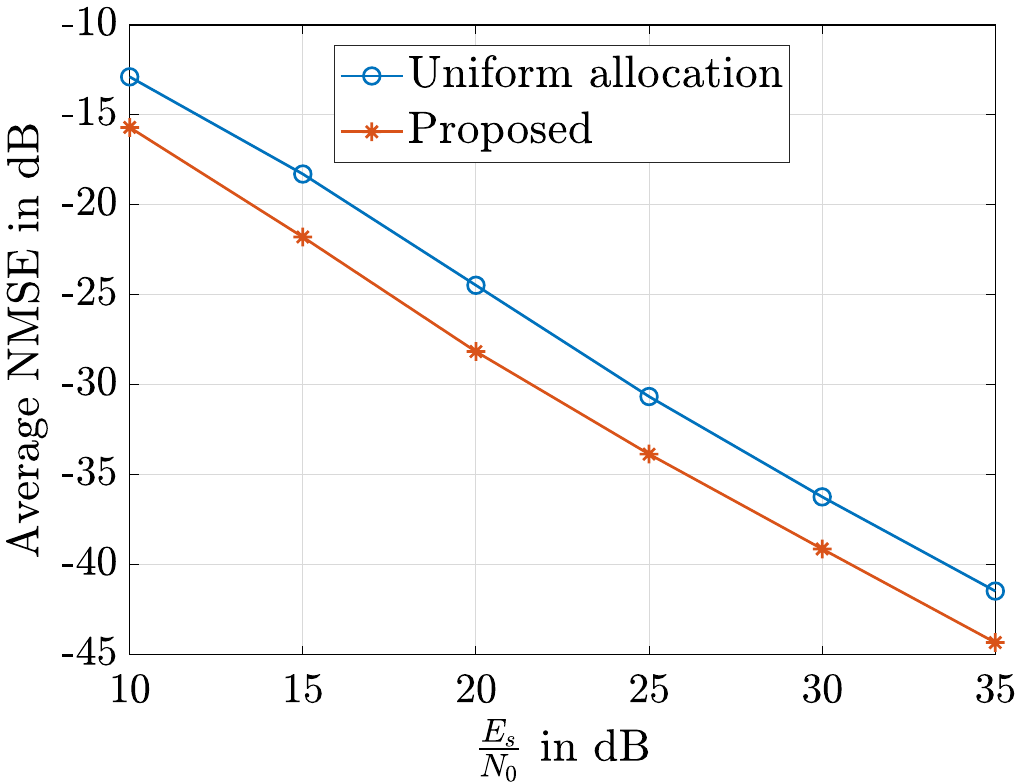}
    \vspace{-4mm} 
\caption{NMSE comparison between uniform and proposed allocations.}
\label{Fig:comparision}
\centering
\vspace{-2mm}
\end{figure}

Next, Fig.~\ref{Fig:comparision} compares the average NMSE performances with different pilot power allocations, i.e., ${\rm{NMSE}}=\frac{1}{K_u}\sum_u^{K_u}{\rm{NMSE}}_{\bm{h}_u}$, in terms of transmission SNR $\frac{E_s}{N_0}$. Note that $E_s=\mathbb{E}[|x[k,l]|^2]=1$ denotes the normalized average data symbol energy and we change $N_0$ to simulate different SNR conditions. We assume the total pilot power $\mathcal{P}=55$ $\text{dB}$. We can observe from Fig.~\ref{Fig:comparision} that the proposed scheme obtains around 3 $\text{dB}$ average NMSE performance improvement than the conventional uniform allocation scheme under the same total pilot power and transmission SNRs.

\section{Conclusion}
In this paper, we proposed a pilot design for multi-user OTFS transmissions based on the derived PCRBs. We first derived the PCRBs in closed form by adopting the \emph{a priori} channel statistical information. Based on the PCRB, an optimization problem for pilot power allocation was formulated and the global optimal allocation is obtained. Numerical results validated our analysis and presented an approximate 3 dB channel estimation NMSE performance improvement, compared to the conventional embedded pilot scheme.

\appendices
\section{Proof of lemma 1}
Recall that $\ln{p(\mathbf{y}|\boldsymbol{\theta},\mathbf{z})}=C-\frac{1}{N_0}\sum_{k,l} \lvert y[k,l]-z[k,l]\rvert^2$, where constant $C$ is independent from $\boldsymbol{\theta}$. For ease of derivation, we omit $[k,l]$ and take 1st-order partial derivative as
\begin{align}
    \frac{\partial \ln  {p(\mathbf{y}|\boldsymbol{\theta},\mathbf{z})} }{\partial\theta_i}&=-\frac{1}{N_0}\sum_{k,l} \left(-\frac{\partial z}{\partial \theta_i}(y^*-z^*)-\frac{\partial z^*}{\partial \theta_i}(y-z)\right).\nonumber
\end{align}
Note that $\mathbb{E}_{y}[y-z]=\mathbb{E}_{y}[y^*-z^*]=0$ because $z$ is the mean of $y$. We can prove the regularity condition of PCRB is satisfied as $\mathbb{E}_{\boldsymbol{\theta}\mathbf{y}}\left[\frac{\partial \ln  {p(\mathbf{y}|\boldsymbol{\theta},\mathbf{z})} }{\partial\theta_i}\right]=0$. Next, we take the second-order partial derivatives as
\begin{align}
    \frac{\partial^2 \ln {p(\mathbf{y}|\boldsymbol{\theta},\mathbf{z})} }{\partial\theta_i\partial\theta_j}=-\frac{1}{N_0}\sum_{k,l} \Big(&-\frac{\partial^2 z}{\partial \theta_i \partial \theta_j}(y^*-z^*)+\frac{\partial z}{\partial \theta_i}\frac{\partial z^*}{\partial\theta_j}\nonumber\\
    &-\frac{\partial^2 z^*}{\partial \theta_i \partial \theta_j}(y-z)+\frac{\partial z^*}{\partial \theta_i}\frac{\partial z}{\partial\theta_j}\Big),\nonumber
\end{align}
where we apply $\mathbb{E}_{y}[y-z]=\mathbb{E}_{y}[y^*-z^*]=0$ again to simplify the result as
\begin{align}
-\mathbb{E}_{\boldsymbol{\theta}\mathbf{y}}\left[\frac{\partial^2 \ln {p(\mathbf{y}|\boldsymbol{\theta},\mathbf{z})} }{\partial\theta_i\partial\theta_j}\right]&=\frac{1}{N_0}\sum_{k,l} \left(\frac{\partial z}{\partial \theta_i}\frac{\partial z^*}{\partial\theta_j}+\frac{\partial z^*}{\partial \theta_i}\frac{\partial z}{\partial\theta_j}\right),\nonumber\\
=&\frac{2}{N_0}\mathbb{E}_{\boldsymbol{\theta}}\left[\Re\left\{\sum_{k,l}\left( \frac{\partial z}{\partial \theta_i}\right)^*\left(\frac{\partial z}{\partial \theta_j}\right)\right\}\right]\label{proveA}
\end{align}

Now, we prove $\mathbf{J}_r$ is a diagonal matrix. Recall that $z[k,l]$ contains $x_u[[k-k_{p,u}]_{N},[l-l_{p,u}]_{M}]$, which is all zeros except the pilot. First, we consider that $\theta_i$ is any parameter of one path and $\theta_j$ is any parameter of any other path. We can observe that equation (\ref{proveA}) contains the component $x_u^*x_u$, which will be zero because the $(k_{p,u},l_{p,u})$ for different paths is different. Next, we consider that $\theta_i$ and $\theta_j$ are any parameters under the same path. If we fix $\theta_i=\Re{(h_{p,u})}$, and $\theta_j=\Im{(h_{p,u})}$, then $\mathbf{J}_{p}^{(i,j)}=\Re{\{j\}}=0$; if $\theta_j=l_{p,u}$ or $\theta_j=k_{p,u}$, then $\mathbf{J}_{p}^{(i,j)}=\mathbb{E}_{\boldsymbol{\theta}}[h_{p,u}]=0$. If we fix $\theta_i=l_{p,u}$ and $\theta_j=k_{p,u}$, then $\mathbf{J}_{p}^{(i,j)}=\mathbb{E}_{\boldsymbol{\theta}}[k_{p,u}]=0$. The remaining off-diagonal elements can be proved to be zero in a similar approach.$\hfill\blacksquare$

\section{Proof of lemma 2}
Recall that $\ln p(\mathbf{h})=\sum_u\sum_p-\frac{{\Re(h_{p,u})^2+\Im(h_{p,u})^2}}{\gamma_{p,u}}-\ln{(\pi \gamma_{p,u})},$ whose first-order partial derivatives are given by
\begin{align}\label{2PK}
    \frac{\partial\ln p(\mathbf{h})}{\partial\theta_i}=
    \begin{cases}
         -\frac{2\Re(h_{p,u})}{\gamma_{p,u}}, &1\le i \le PK_u,\\
         -\frac{2\Im(h_{p,u})}{\gamma_{p,u}}, &PK_u< i \le 2PK_u,
    \end{cases}
\end{align}
and for $2PK_u< i \le 4PK_u$, we have
 
\begin{align}\label{4PK}
    \frac{\partial\ln p(\mathbf{h})}{\partial\theta_i}=-\frac{1}{\gamma_{p,u}}\frac{\partial\gamma_{p,u}}{\partial\theta_i}+\frac{|h_{p,u}|^2}{\gamma_{p,u}^2}\frac{\partial\gamma_{p,u}}{\partial\theta_i}.
\end{align}
Note that we can prove the regularity condition by substituting $\mathbb{E}[h_{p,u}]=0$ and $\mathbb{E}[|h_{p,u}|^2]=\gamma_{p,u}$ into (\ref{2PK}) and (\ref{4PK}), respectively. After taking the second-order partial derivatives and expectations, we can obtain
\begin{align}
\mathbf{J}_p^{(i,i)}=\mathbb{E}_{\boldsymbol{\theta}}\left[\frac{2}{\gamma_{p,u}}\right], &1\le i \le 2PK_u.
\end{align}
Note that we obtain the results for $2PK_u< i,j \le 4PK_u$ in a similar approach by using $\mathbb{E}[h_{p,u}]=0$, $\mathbb{E}[|h_{p,u}|^2]=\gamma_{p,u}$, and $\mathbb{E}[k_{p,u}]=0$ suitably, which are omitted due to the space limitations of the paper. Finally, we give hints on proving the diagonal property of $\mathbf{J}_p$. If $\theta_i=\Re{(h_{p,u})}$, then $\mathbf{J}_{r}^{(i,j)}=\mathbb{E}_{\boldsymbol{\theta}}[\Re{\{h_{p,u}\}}]=0, \forall i\neq j$. If we fix $\theta_i=k_{p,u}$, then $\mathbf{J}_{r}^{(i,j)}=0, \forall i\neq j$ because $p(\mathbf{h})$ is independent of $k_{p,u}$.
$\hfill\blacksquare$
\bibliographystyle{IEEEtran}
\bibliography{ref}
\end{document}